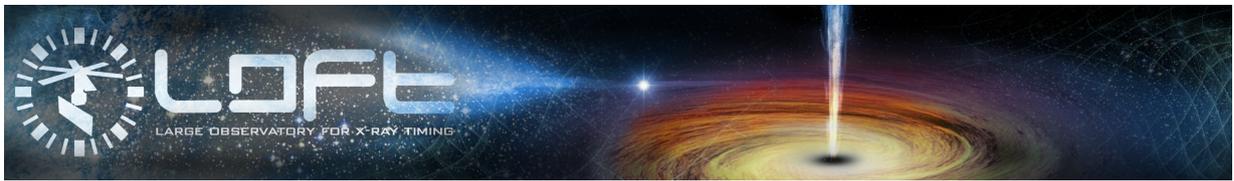

# High-energy radiation from thunderstorms and lightning with *LOFT*

## White Paper in Support of the Mission Concept of the Large Observatory for X-ray Timing

### Authors

M. Marisaldi[1,2], D.M. Smith[3], S. Brandt[4], M.S. Briggs[5], C. Budtz-Jørgensen[4], R. Campana[1], B.E. Carlson[6], S. Celestin[7], V. Connaughton[5], S.A. Cummer[8], J.R. Dwyer[9], G.J. Fishman[10], M. Fullekrug[11], F. Fuschino[12,1], T. Gjesteland[2], T. Neubert[4], N. Østgaard[2], M. Tavani[13]

[1] National Institute for Astrophysics (INAF) - IASF Bologna, Bologna, Italy.

[2] Birkeland Centre for Space Science, University of Bergen, Bergen, Norway.

[3] Santa Cruz Institute for Particle Physics, Physics Department, University of California, Santa Cruz, California, USA.

[4] Technical University of Denmark, National Space Institute (DTU Space), Kgs. Lyngby, Denmark.

[5] Center for Space Physics and Aeronomy, University of Alabama in Huntsville, Huntsville, Alabama, USA.

[6] Carthage College, Wisconsin, USA.

[7] University of Orleans, LPC2E, CNRS, France.

[8] Electrical and Computer Engineering Department, Duke University, Durham, NC, USA

[9] Univerity of New Hampshire, Durham, NH, USA.

[10] Space Science Office, NASA Marshall Space Flight Center, Huntsville, Alabama, USA.

[11] University of Bath, United Kingdom.

[12] Department of Physics and Astronomy, University of Bologna, Bologna, Italy.

[13] National Institute for Astrophysics (INAF) - IAPS, Roma, Italy.

## Preamble

The Large Observatory for X-ray Timing, *LOFT*, is designed to perform fast X-ray timing and spectroscopy with uniquely large throughput (Feroci et al., 2014). *LOFT* focuses on two fundamental questions of ESA's Cosmic Vision Theme "Matter under extreme conditions": what is the equation of state of ultra-dense matter in neutron stars? Does matter orbiting close to the event horizon follow the predictions of general relativity? These goals are elaborated in the mission Yellow Book (`http://sci.esa.int/loft/53447-loft-yellow-book/`) describing the *LOFT* mission as proposed in M3, which closely resembles the *LOFT* mission now being proposed for M4.

The extensive assessment study of *LOFT* as ESA's M3 mission candidate demonstrates the high level of maturity and the technical feasibility of the mission, as well as the scientific importance of its unique core science goals. For this reason, the *LOFT* development has been continued, aiming at the new M4 launch opportunity, for which the M3 science goals have been confirmed. The unprecedentedly large effective area, large grasp, and spectroscopic capabilities of *LOFT*'s instruments make the mission capable of state-of-the-art science not only for its core science case, but also for many other open questions in astrophysics.

*LOFT*'s primary instrument is the Large Area Detector (LAD), a $8.5 \, \text{m}^2$ instrument operating in the 2–30 keV energy range, which will revolutionise studies of Galactic and extragalactic X-ray sources down to their fundamental time scales. The mission also features a Wide Field Monitor (WFM), which in the 2–50 keV range simultaneously observes more than a third of the sky at any time, detecting objects down to mCrab fluxes and providing data with excellent timing and spectral resolution. Additionally, the mission is equipped with an on-board alert system for the detection and rapid broadcasting to the ground of celestial bright and fast outbursts of X-rays (particularly, Gamma-ray Bursts).

This paper is one of twelve White Papers that illustrate the unique potential of *LOFT* as an X-ray observatory in a variety of astrophysical fields in addition to the core science.

# 1 Summary

In this paper we explore the possible contributions of the Large Observatory for X-ray Timing (*LOFT*) mission to the study of high-energy radiation from thunderstorms and lightning. *LOFT* is a mission dedicated to X-ray timing studies of astrophysical sources, characterised by a very large effective area of about 8.5 square meters at 8 keV. Although the main scientific target of the mission is the fundamental physics of matter under extreme conditions, the peculiar instrument concept allows significant contributions to a wide range of other science topics, including the cross-disciplinary field of high-energy atmospheric physics, at the crossroad between geophysics, space physics and astrophysics. In this field we foresee the following major contributions:

- detect $\approx 700$ Terrestrial Gamma-ray Flashes (TGFs) per year, probing the TGF intensity distribution at low fluence values and providing an unbiased sample of bright events thanks to the intrinsic robustness against dead-time and pile-up;

- provide the largest TGF detection rate surface density above the equator, allowing for correlation studies with lightning activity on short time scales and small regional scales, to probe the TGF / lightning relationship;

- lower by an order of magnitude the minimum detectable fluence for Terrestrial Electron Beams (TEBs), an additional tool to probe TGF production mechanism and the lower edge of TGF intensity distribution;

- open up a discovery space for the detection of high-altitude electron beams and weak X-ray emissions associated to Transient Luminous Events (TLEs).

# 2 Introduction

Earth's atmosphere is a complex system still far from being understood in its entirety. The same lightning phenomena, so pervasive and familiar to everyone, present several unsolved problems, especially concerning the processes of lightning initiation and propagation and the distribution of the available energy budget between electromagnetic, thermal and mechanical processes (Dwyer & Uman, 2014). Moreover, lightnings in the troposphere are closely related to Transient Luminous Events (TLEs), a wide range of optical phenomena that take place in the stratosphere, mesosphere, up to the lower thermosphere (Pasko et al., 2012). In addition, thunderstorms are observed to emit energetic particles and photons on several time scales, spanning from quasi-stationary gamma-ray glows, to tens of millisecond long bursts of electrons, and sub-millisecond bright gamma-ray flashes. In fact, thunderstorms are now established as the most energetic natural particle accelerators on Earth, capable of accelerating electrons up to several tens of MeV, see Dwyer et al. (2012) for a recent comprehensive review.

The physical mechanism at the basis of such energetic radiation is now widely accepted to be the production of runaway electrons in thunderstorms electric field, capable to accelerate free electrons to relativistic energies when the net energy gain due to the ambient electric field is larger than the energy loss by collisions with air molecules. These collisions may lead to the production of new seed electrons in the runaway regime, driving an exponential avalanche multiplication (Gurevich et al., 1992) possibly enhanced by a feedback mechanism ruled by backward propagating secondary photons and positrons, the Relativistic Feedback (RF) (Dwyer, 2012). This exponentially growing population of energetic electrons can then produce Bremsstrahlung photons by interaction with air molecules resulting in bright flashes of gamma-rays detected at satellite altitudes, very far from the production region, known as Terrestrial Gamma-ray Flashes (TGF). Discovered serendipitously in the early nineties by the BATSE detector onboard the Compton Gamma-Ray Observatory (Fishman et al., 1994), TGFs are now currently observed by detectors onboard three space missions devoted to X and gamma-ray astrophysics, *RHESSI* (Smith et al., 2005), *AGILE* (Marisaldi et al., 2014) and *Fermi* (Briggs et al., 2013). As a byproduct of TGF photons interacting in the upper atmospheric layers, also pulses of energetic electrons and positrons known as Terrestrial Electron Beams (TEBs) can be detected at satellite altitude (Dwyer et al., 2008; Briggs et al., 2011). These elusive and poorly characterized events are shaped by the geomagnetic field along which particles have traveled, and carry information about their parent TGFs.

Despite a general consensus on the underlying physical production mechanism, several issues are still not clear, most notably the link between TGFs and lightning initiation and propagation, and whether production takes place in the macroscopic thundercloud electric field, or in the local strong and variable electric field at lightning leader tip during discharge upward propagation (Celestin & Pasko, 2011). Probably, the most compelling question about TGFs and related phenomena is whether these events are relatively rare and linked to a specific class of lightning, or pervasive and possibly associated to every lightning discharge, as suggested by some recent studies (Østgaard et al., 2012). The answer to this question is relevant to assess the overall impact of the energy delivered as high-energy radiation in the atmosphere system as a whole, including the coupling and feedback between the troposphere and the upper atmospheric layers. To address these questions, current observations are still sparse and hampered by instrumental issues such as dead-time and pile-up. Future missions such as *ASIM* (Neubert, 2009) and *TARANIS* (Lefeuvre et al., 2009), which are specifically designed to study these phenomena, will be operative during years 2016–2020 and provide a large wealth of observations to shed new light on these phenomena. However, the physical geometrical area of the proposed detectors is still of order of $\approx 1000 \, \text{cm}^2$.

In this paper we explore the possible contributions to this field by the Large Observatory for x-ray Timing (*LOFT*) mission (Feroci et al., 2012). *LOFT*'s Large Area Detector (LAD) exhibits an overwhelmingly large effective area in the X-ray band 2–30 keV, which is out of reach for all past and currently planned space missions. Although designed for timing studies of astrophysical sources, *LOFT* has a wide range of other science objectives, and we believe this large effective area can be fruitful in the study of TGFs and related phenomena. Moreover, the LAD is composed of ~360000 independent readout channels, each of them corresponding to about 34 mm$^2$ sensitive area, therefore providing an exceptional robustness against dead-time and pile-up, which strongly affect current TGF observations. In addition, *LOFT* observation mode foresees data download on a photon-by-photon basis, recording timing and energy information of all detected counts, therefore providing the most detailed picture of the detected events.

Figure 1 shows the *LOFT* LAD nominal on-axis effective area compared to that of current and future TGF detectors. Since TGFs and related phenomena have an energy spectrum peaking in the MeV range, the LAD energy range is not optimized for their detection. This translates into a significant reduction of sensitivity when convolution with the appropriate source spectrum is considered. Still, the effective area remains relevant compared to current and future missions, given the very large initial geometrical area. Figure 2 shows the *LOFT* LAD total effective area as a function of the off-axis angle for a TGF and Terrestrial Electron Beam (TEB) spectra, for different energy ranges.

Here we consider both the nominal 2–30 keV and the extended 2–80 keV energy ranges. In addition we explore also the capabilities of an observing mode capable to downlink the time stamps of all counts, including overflow events which are rejected at the level of the back-end electronics in the current design. Such an observing mode strongly enhances the total effective area, at the price of a larger background rate, and can be considered as a dedicated observing mode to be enabled during Earth occultation under certain conditions only.

TGF and TEB spectra provided in Dwyer & Smith (2005) and Dwyer et al. (2008) were simulated with GEANT4 and propagated through the full LAD mass model (Campana et al., 2013) at different off-axis angles in order to obtain the average detection efficiency. This detection efficiency was then multiplied with the LAD geometrical area projected along the direction of incidence to obtain the cumulative effective area for a given spectrum. The use of Monte Carlo simulations was mandatory since the source spectrum is mostly outside the LAD nominal range and, in that range, the response of the instrument is strongly non-diagonal, i.e., interacting photons or electrons release in the detector a small portion of their full energy.

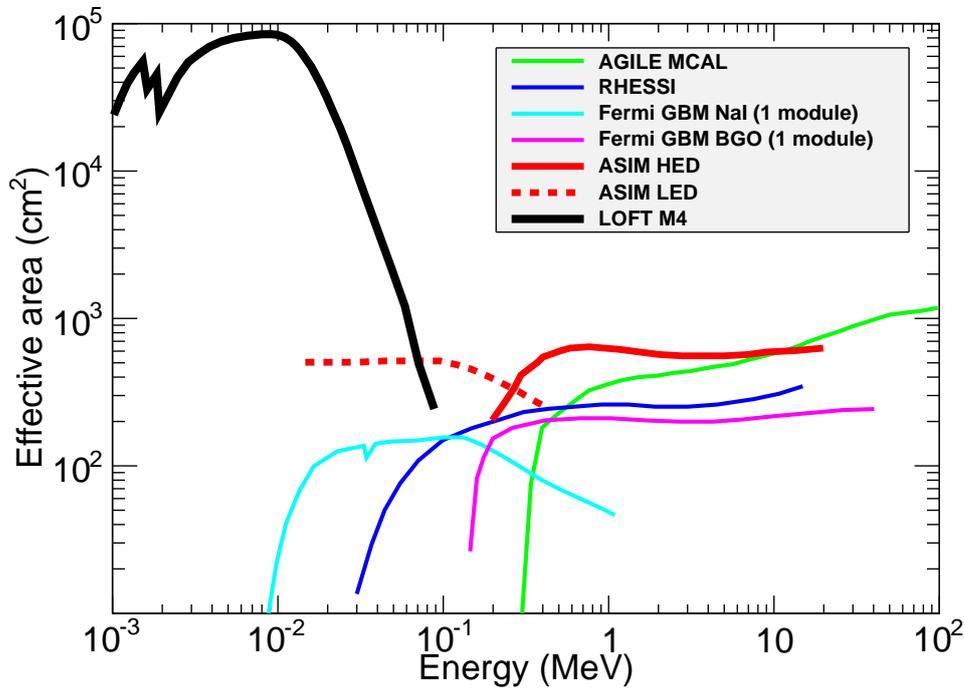

Figure 1: Effective area of *LOFT* LAD and current and future TGF detectors.

In the following we discuss the relevant contributions that *LOFT* can give to assess the lower and upper edges of the intensity distributions of TGFs (Sect. 3) and associated TEBs (Sect. 4), then we speculate on the possible discovery space opened by *LOFT* performance for the detection of energetic radiation associated to transient luminous events (Sect. 5).

## 3   Exploring the edges of TGF intensity distribution

The TGF intensity distribution is observed to follow a power law behavior with a $-2.3 \pm 0.1$ spectral index. This result has been remarkably confirmed by independent analysis on all currently active TGF detectors (Østgaard et al., 2012; Tierney et al., 2013; Marisaldi et al., 2014). In particular, Tierney et al. (2013) showed that there is no evidence for a break or a roll-off at low intensity values, suggesting the possibility that current instruments are just detecting the tip of the iceberg and TGF intensity distribution may extend unaltered to very low fluence values. Addressing the low edge of the TGF intensity distribution is therefore an important issue to understand the overall relationship between TGF and lightning, and ultimately assess whether TGFs are rare or rather pervasive phenomena. Extrapolation of the intensity distribution as currently known is compatible with the hypothesis that every lightning is associated to a TGF (Østgaard et al., 2012), which in turn has strong implications for production models. However, there is a big gap to bridge between currently observed TGF/lightning flash ratio of $\sim 10^{-4}$ (Fuschino et al., 2011; Briggs et al., 2013) and $\approx 1$.

Table 1 shows average effective area, expected background level, minimum detectable counts and minimum detectable TGF fluence for some energy range and *LOFT* observing modes. Background level has been carefully estimated by simulating all the contributions expected for a mission in Low-Earth Orbit (Campana et al., 2013). The background in the 2–80 keV energy range is largely dominated by the cosmic x-ray background. If we consider a 300 $\mu$s time window for TGF search, which is a reasonable time interval according to measured TGF duration distribution (Briggs et al., 2013), and assuming a maximum acceptable rate of one false event per year due to statistical fluctuations (Poisson probability $< 9.5 \times 10^{-12}$), we can estimate the minimum number of counts in the search window for the different energy ranges, according to the expected background. We can then obtain the minimum detectable TGF fluence by dividing the number of counts by the cumulative effective area. The calculations have been performed for two observing conditions: during Earth occultation (no source within the LAD field of view (FOV)), and during nominal observations of a 100 mCrab source, which represents a typical astrophysical target of the *LOFT* mission. Such a source, which is the main signal for the *LOFT* mission core science, in fact acts as background for TGF observations, delivering $\sim 20 \, \text{kcounts s}^{-1}$, mostly in the 2–30 keV energy range. Therefore TGF minimum detectable fluence is larger during observations than during occultation. According to the *LOFT* mission science requirements, we can safely assume that 70% of the time will be spent in observation and 30% in Earth occultation.

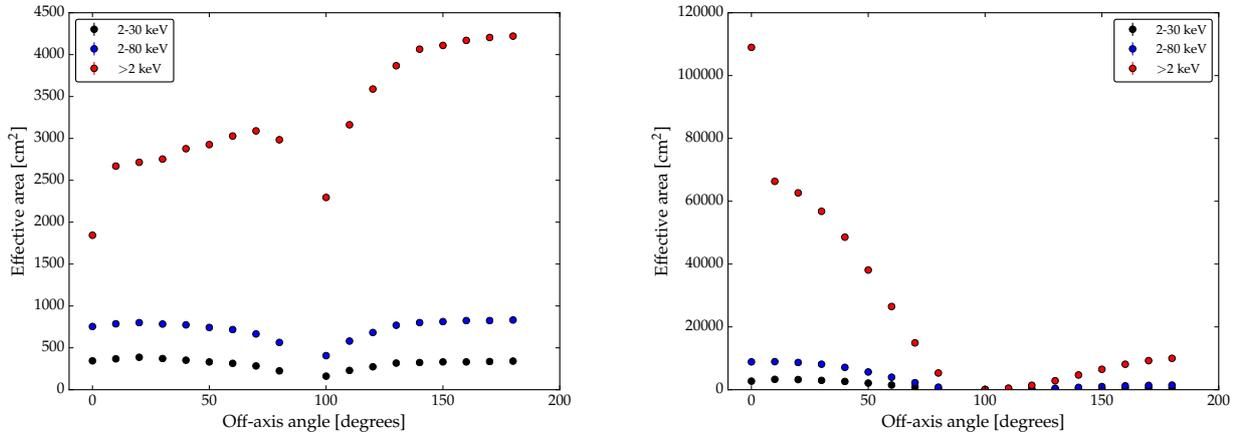

Figure 2: *Left: LOFT* LAD cumulative effective area for a typical TGF spectrum and *right:* electron beam spectrum as a function of the off-axis angle, for three different energy ranges.

*RHESSI* TGF detection rate is 0.42 TGFs/day (Grefenstette et al., 2009; Østgaard et al., 2012). Considering the trigger conditions reported in (Grefenstette et al., 2009) (17 counts, two of which of background, on average) and an average effective area of 256 cm$^2$, the minimum TGF fluence at *RHESSI* detection threshold is 0.058 cm$^{-2}$. We can then estimate the expected TGF detection rate on a *RHESSI*-like orbit (inclination 35°) for a lower fluence threshold by extrapolating the observed $-2.3$ index power law intensity distribution. The minimum detectable fluence of 0.019 cm$^{-2}$ estimated for *LOFT* during Earth occultation in the 2–80 keV energy range, results in a factor of four improvement in detection rate with respect to *RHESSI*.

Considering that TGF latitudinal distribution peaks close to the equator (Grefenstette et al., 2009), we can estimate an additional factor 2.2 of improvement in detection rate with respect to *RHESSI* due to *LOFT* very low inclination equatorial orbit. Given this orbital factor and the expected duty cycle between observation and Earth occultation, we can estimate a detection rate for *LOFT* of $\sim 2 \, \text{TGFs day}^{-1}$.

Although significant, such a TGF detection rate is not a compelling result *per se*, since a similar detection rate is already achieved by the *Fermi* mission operating in continuous event mode (Briggs et al., 2013). However, all *LOFT* TGFs will be detected in a narrow equatorial belt, given the low inclination orbit expected. This will provide the largest TGF detection rate surface density, i.e., the number of detected TGFs per unit time per square degree, ever achieved. This is an important parameter for time-dependent correlative studies between TGFs and lighting activity at regional scales. All three currently active TGF detectors have shown still unexplained regional variations in the TGF/lightning flash ratio (Smith et al., 2010; Fuschino et al., 2011; Briggs et al., 2013). All these results are based on lightning activity maps averaged over several years (Christian et al., 2003), because the current instruments detection rate surface density does not allow one-to-one comparison with lightning maps on year or seasonal time scales. This latter result shall be accomplished by *LOFT*, providing further insight on the TGF lightning relation. Moreover, *LOFT* will be operative together with the next generation of geostationary lightning imagers onboard *GOES-R* and Meteosat Third Generation (MTG) satellites. These detailed observations at low latitude will be complementary to the high-latitude TGF observations provided by the forthcoming *ASIM* and *TARANIS* missions. We also note that this rate is based on simulations of a TGF spectrum well fitting the *RHESSI* cumulative spectrum (Dwyer & Smith, 2005). We know that TGF spectra should get softer as the distance between the source and the sub-satellite point increases (Østgaard et al., 2008; Cramer et al., 2012), therefore the detection of soft events not detected by current instruments could be enhanced for *LOFT*, resulting in a larger detection rate than estimated.

Table 1: Expected background rate $N_{bkg}$, cumulative effective area $A^*_{eff}$, minimum number of counts at trigger threshold $N^*_{min}$ (0.3 ms and 10 ms time window for TGFs and TEBs, respectively), minimum detectable fluence $F^*_{min}$ for both TGFs and TEBs, for different *LOFT* LAD energy range configurations and orbital phases.

| Orbital phase | E range (keV) | $N_{bkg}$ ($s^{-1}$) | TGFs $A^{TGF}_{eff}$ ($cm^2$) | $N^{TGF}_{min}$ | $F^{TGF}_{min}$ ($cm^{-2}$) | TEBs $A^{TEB}_{eff}$ ($cm^2$) | $N^{TEB}_{min}$ | $F^{TEB}_{min}$ ($cm^{-2}$) |
|---|---|---|---|---|---|---|---|---|
| Earth | 2–30 | 1400 | 270 | 10 | 0.037 | 860 | 43 | 0.050 |
| occultation | 2–80 | 2500 | 630 | 12 | 0.019 | 2300 | 62 | 0.027 |
| | >2 | 60000 | 2900 | 53 | 0.018 | 16000 | 757 | 0.047 |
| | 2–30 | 21400 | 260 | 30 | 0.120 | 860 | 310 | 0.360 |
| Observation | 2–80 | 22500 | 630 | 30 | 0.048 | 2300 | 324 | 0.141 |
| | >2 | 80000 | 3300 | 64 | 0.019 | 16000 | 981 | 0.061 |

On the other edge of the TGF intensity distribution, observation of bright events by all currently operating detectors are hampered by dead-time and pile-up effects (Gjesteland et al., 2010; Briggs et al., 2010; Østgaard et al., 2012; Marisaldi et al., 2014). Therefore, all discussion of bright events are based on detailed modeling of the detectors and associated electronics under high count-rate conditions. This will be partially overcome by the forthcoming ASIM and TARANIS missions by a suitable detector design. The *LOFT* LAD extremely high degree of spatial segmentation (~36000 readout channels, each corresponding to about 34 mm$^2$ sensitive area) will provide an intrinsic robustness against dead-time and pile-up effects, allowing for the unbiased reconstruction of the time profile of bright events with a time resolution limited by counting statistics and by the detector timing accuracy ($10\,\mu s$) only.

## 4  Mapping terrestrial electron beams

Since TGFs are the result of Bremsstrahlung emission of a population of energetic electrons propagating through the atmosphere, soon after their discovery it was speculated whether the electron themselves could be detected at satellite altitude. After TGFs were firmly associated to thundercloud tops (Dwyer & Smith, 2005), it was realised that primary electrons could not travel at satellite altitudes because of atmospheric absorption. However, TGF photons can still interact by Compton scattering and pair production with air molecules above the production region producing a population of secondary energetic electrons and positrons which can travel at satellite altitude because of the lower atmospheric absorption, resulting in a so-called Terrestrial Electron Beam (TEB). Although the overall number of these secondary electrons is a small fraction of the initial number of photons, they travel along the geomagnetic field lines remaining confined in a small region, resulting in several peculiar properties. First of all, confinement makes the flux almost constant with distance, whereas TGF photon flux scales down as the inverse square of the distance from the source. However, for the same reason the spatial extent of a TEB at satellite altitudes is much smaller than that of a TGF. The net result of confinement is that the observed fluence of electron beams at satellite altitude is comparable to that of TGFs, but the observation frequency is much smaller.

Moreover, TEBs seen at the magnetic conjugate point from their origin, when Earth's field is stronger on the satellite end of the field line, display a unique temporal signature. Electrons with large pitch angles mirror below the satellite but above the atmosphere and come back, producing in detectors two count peaks separated by a few tens of milliseconds, depending on the position of the satellite and TGF along the geomagnetic field line, the second peak being dimmer than the first one because of atmospheric loss of the particles at small pitch angles. This peculiar feature, plus the typical asymmetric pulse profile due to electron velocity dispersion, make TEB easily recognisable in time series data, reducing the probability of false triggers due to background fluctuations and hence lowering the minimum detectable fluence.

Dwyer et al. (2008) showed that some of the BATSE long-duration, multi-peaked events initially classified as TGFs were actually most probably TEBs, suggesting also the production mechanism. TEBs with magnetically mirrored secondary peak were also detected in *RHESSI* and *Fermi* data (Briggs et al., 2011), which provided the largest sample of TEBs to date: 6 TEBs and 77 TGFs detected during two years under the same trigger conditions. This detection rate ratio is relatively large compared to expectations based solely on the relative ratio between TGFs and TEBs surface extent at satellite altitude, therefore Carlson et al. (2011) suggested that many detected TEBs are the signature of a population of weak TGFs not detectable in gamma-rays because of the low fluence at satellite altitude. Interestingly, the detection of TEBs is therefore a means to further dig into the lower edge of the TGF intensity distribution, unfortunately hampered by the intrinsic observation difficulty due to the narrow beam geometry.

The improved detection method implemented for *Fermi* described by Briggs et al. (2013) strongly enhanced the TGF detection rate by improving sensitivity for very short time scales, but barely changed TEB detection rate, reconciling the TEB / TGF ratio with expectations. Moreover, *Fermi* results show that the current rate is mostly the maximum that can be expected given the detector characteristics and effective area. Even future missions such as *TARANIS* (Lefeuvre et al., 2008), that will have dedicated photon and electron discrimination capability, will be limited to a geometrical area lower than $1000\,cm^2$, therefore providing only a moderate improvement in TEB detection rate and minimum detectable fluence. *LOFT* LAD effective area for TEBs, averaged over the full solid angle, is about $16000\,cm^2$ if the full energy range above $2\,keV$ is considered, two orders of magnitude larger than the $\simeq 160\,cm^2$ *Fermi* reported for the GBM BGO detector (Meegan et al., 2009; Tierney et al., 2013). The minimum detectable TEB fluence in this energy range is however limited by a large background rate, so that the best results in terms of sensitivity are obtained, like for TGFs, in the energy range 2–80 keV. In this range the minimum detectable fluence is $0.027\,cm^{-2}$, a factor $\approx 5$ lower than that of *Fermi*. We note that considering an effective area averaged over the full solid angle is particularly appropriate for TEBs since any event will include electrons coming from many different angles, and with angular distribution evolving with time. Assuming that the TEB fluence scales as the parent TGF fluence distribution discussed in Sect. 3, this improvement in the minimum detectable fluence would result in a factor $\approx 8$ improvement in detection rate. However, it is now difficult to estimate the expected detection rate for *LOFT* because a simple scaling based on the orbital inclination, as previously done for TGFs, is not appropriate. In fact, TEBs follow the geomagnetic field lines in a complex fashion, as pointed out by Xiong et al. (2012), resulting in significant detection probabilities even at high latitudes and a strong dependence on satellite altitude. We note however that some of the most active regions for TEB detection, in particular central America and equatorial Africa, will be observable by *LOFT*.

## 5 Discovery space

*LOFT*'s combined high efficiency at low energies and huge collecting area combine to create a large discovery space for terrestrial X-ray events with softer spectra than TGFs. Three categories of transient luminous events (TLE) associated with thunderstorms occur high enough in the atmosphere for X-rays of 10s of keV to escape, and in each case there is at least a qualitative justification for expecting some X-ray or energetic electron emission. Sprites (Sentman et al., 1995) are streamer breakdowns initiating around 70 km that occur when the electric field at those altitudes is suddenly driven above the streamer threshold by the redistribution of charge in a large storm system below by lightning. Models of sprite streamer tips show that they are capable of producing energetic electrons (Chanrion & Neubert, 2010), some of which could either escape directly to space or create bremsstrahlung. Elves are expanding rings of light at even higher altitude caused by acceleration of electrons in the lower ionosphere by the electromagnetic pulse (EMP) of high-peak-current lightning. The same mechanism was once proposed to explain TGFs themselves (Inan & Lehtinen, 2005). This model is no longer favored for TGFs for a number of reasons, but might remain viable as a source of fainter, softer emission to which *LOFT* could be sensitive. Finally, gigantic jets (GJs) (Su et al., 2003) are thought to be lightning leaders diverted to the ionosphere instead of terminating in a cloud charge center or at the ground. The leader will transition to purely streamer propagation at an altitude that depends on the potential it carries (Silva & Pasko, 2013), and either the leader or streamer phases could serve as sources of X-rays that *LOFT* could detect: the leader by the same process that produces X-rays from stepped leaders near the ground (Moore et al., 2001), and the streamers by the same mechanism hypothesized for sprites.

## 6  Conclusions

We presented here the capabilities of the *LOFT* mission, specifically designed for X-ray timing studies of astrophysical sources, for the study of high-energy radiation emission from thunderstorms and lightning. We expect the detection of $\approx 700$ TGFs per year across the equator, providing the largest detection rate surface density ever achieved. This parameter is important for correlation studies of TGFs and global lighting activity detected from space at short time scales, to shed new light on the TGF lightning relation. This capability is strategic in view of the next generation lightning imagers onboard the geostationary *GOES-R* and Meteosat Third Generation satellites, that will be operative during *LOFT* lifetime. Concerning TEBs, we expect an order of magnitude improvement in the minimum detectable fluence with respect to current missions, resulting in a larger dataset mandatory to study this still poorly characterized phenomenon. Moreover, although not specifically designed for TGF observation, the high degree of spatial segmentation of the *LOFT* detector will prevent dead-time and pile-up issues that hamper observation of bright events by current missions, providing an unbiased sample of very bright events. This is all guaranteed science that can be expected based on available literature. In addition, we must note that *LOFT* will open an observational window for dim X-rays or particle transients of atmospheric origin never explored before, potentially resulting in new discovery space.